\documentclass[twocolumn,aps,10pt,showkeys,showpacs,preprintnumbers,prd,superscriptaddress,nofootinbib]{revtex4-1}
\usepackage{float}
\usepackage{graphicx}	
\usepackage{amssymb}
\usepackage{dcolumn}
\usepackage{mathtools}
\usepackage{amsmath}
\usepackage{xcolor}
\usepackage{color}
\usepackage[normalem]{ulem}
\usepackage{theorem}
\usepackage{subfigure, rotating, bm, array}
\usepackage[pagebackref=false, colorlinks=true]{hyperref}
\hypersetup{linkcolor=blue, citecolor=blue,urlcolor=blue}

\begin{document}

\title{Quasinormal ringing and Unruh-Verlinde Temperature of the Frolov Black Hole}

\author{Akshat Pathrikar}
\email{akshatpathrikar014@gmail.com}
\affiliation{International Centre for Space and Cosmology, Ahmedabad University, Ahmedabad 380009, Gujarat, India}


\begin{abstract}
\noindent
In this study, we investigate electromagnetic and Dirac test field perturbations of a charged regular black hole arising from quantum gravity effects, commonly referred to as the Frolov black hole, a regular (nonsingular) black hole solution. We derive the master wave equations for massless electromagnetic and Dirac perturbations and solve them using the standard Wentzel-Kramers-Brillouin (WKB) method along with Padé Averaging. From these solutions, we extract the dominant and overtone quasinormal mode (QNM) frequencies along with the associated grey-body factors, highlighting the deviations introduced by quantum gravity corrections compared to the classical case of Reissner–Nordström black hole. Furthermore, we analyze the Unruh-Verlinde temperature of this spacetime, providing quantitative estimates of how quantum gravity effects influence both quasinormal ringing and particle emission in nonsingular black hole models.

\vspace{.6cm}
$\boldsymbol{Key words}$ : Quasinormal modes, Regular Black Holes, Unruh Temperature, WKB method.
\end{abstract}

\maketitle

\section{Introduction}
\label{sec:intro}

The General Theory of Relativity (GR), our modern understanding of gravitation, has enjoyed remarkable success for over a century, passing numerous experimental and observational tests with great precision. Nevertheless, under certain extreme conditions, GR predicts its own breakdown. For instance, the celebrated singularity theorems of R. Penrose and S.W. Hawking \cite{Penrose:1964wq, Hawking:1966vg} demonstrate that spacetime singularities are inevitable within black holes. In these regimes, quantum effects cannot be neglected, signaling that classical GR is no longer sufficient. It is also known that at high densities of matter, quantum effects become important, and the matter pressure
may be able to counterbalance gravitational collapse, and it seems reasonable that when
matter reaches Planck density, which is the onset of quantum gravity effects, and there would be
enough pressure as to prevent the formation of a singularity. This situation has motivated
the derivation and study of non-singular or regular black holes.\\

The question of how general relativity might avoid the formation of spacetime singularities is both long standing and of more than purely formal interest. In 1968, Bardeen presented the first example of a regular black hole—a spacetime with an event horizon but free of singularities, while still obeying the weak energy conditions \cite{refId0}. Although conceptually important, Bardeen’s solution lacked for many years a clear physical interpretation because it is not a vacuum solution of Einstein’s equations. To obtain it, one must either introduce some form of external matter or consider modifications to gravity. In fact, Bardeen achieved this geometry by positing an ad hoc stress–energy tensor that is finite everywhere, falls off at infinity, and satisfies the weak energy conditions.
Bardeen’s construction inspired numerous subsequent models exploring the mechanisms by which singularities might be avoided. Several alternative regular black hole solutions have since appeared in the literature, particularly within theories where gravity couples to nonlinear electrodynamics \cite{Ayon-Beato:2000mjt, Dymnikova:2004zc}. Notably, Dymnikova proposed a black hole whose interior is described by a de Sitter core smoothly matching a Schwarzschild exterior \cite{Dymnikova1992}. Further developments and variations of regular black-hole spacetimes have continued to be investigated in many works \cite{Ansoldi:2008jw, Ovalle:2024wtv, Ovalle:2025pue, Lemos:2011dq}.

Regular black holes can be broadly grouped according to the geometry near their center: 
those featuring a de Sitter (dS) core and those with a Minkowski-like core. 
Well-known dS-core solutions include the Bardeen black hole, 
the Hayward black hole~\cite{Hayward:2005gi}, and the Frolov black hole~\cite{Frolov:2016pav}.  
In contrast, regular black holes with a Minkowskian interior are typically described 
using exponential-type potentials, as discussed in Refs.~\cite{Xiang:2013sza, Culetu:2013fsa, Culetu:2014lca, Rodrigues:2015ayd, Simpson:2019mud, Ghosh:2014pba, Li:2016yfd, Martinis:2010zk, Ling:2021olm}.  

V.~L.~Frolov constructed several metrics under a set of intuitive assumptions 
aimed at finding non-singular black-hole geometries without altering general relativity itself. 
Extending his analysis to the charged case, Frolov introduced variants such as a modified Hayward solution.  
In his framework, a characteristic length parameter $\alpha_{0}$ (denoted $l$ in the original paper) 
is tied to a critical energy scale $\mu$ through $\alpha_{0} = \mu^{-1}$. 
Thus, alongside the black hole mass, the parameter $\alpha_{0}$ sets the scale at which departures 
from Einstein’s equations become important.  
More precisely, this scale is reached when $\alpha_{0}^{-2}$ is comparable to the curvature scalar $R$.  
Frolov also argued that one may continue to employ the usual metric tensor $g_{\mu\nu}$, 
while acknowledging a separate quantum-gravity length $\lambda$, 
significantly smaller than $\alpha_{0}$, where such effects dominate.  
These considerations ensure that the resulting spacetime remains regular even at $r = 0$.\\

In this work, our goal is to investigate some of the fundamental observable characteristics of quantum-corrected black holes. One of the most important such characteristics is the spectrum of quasinormal modes, these are the damped oscillations that dominate the ringdown phase of a perturbed spacetime at intermediate to late times \cite{Konoplya:2011qq, Berti:2009kk}. These frequencies are often referred to as the “fingerprints” of a black hole, as they are independent of the specific perturbation that excites them and depend entirely on the underlying geometry of the spacetime. Another essential quantity we study is the set of grey-body factors, which determine the fraction of Hawking radiation that can tunnel through the black hole’s effective potential barrier and propagate to infinity, which might be detected by a distant observer \cite{Konoplya:2024lir, Tang:2025mkk, Bolokhov:2024otn}. We also study the Unruh temperature, which arises due to the response of an accelerating observer in a vacuum. An observer undergoing constant acceleration perceives the vacuum as a thermal bath of particles with a characteristic temperature and in the context of black holes, the Unruh temperature serves as a local notion of temperature near the horizon proportional to their acceleration \cite{Dubinsky:2025fwv, Konoplya:2021ube}. These quantities provide critical information about the near-horizon geometry and the nature of quantum corrections. Besides, from an observational perspective, most current experimental data arise from the detection of gravitational waves emitted by coalescing black holes and from electromagnetic observations of their surrounding environments. However, both approaches still leave a wide parameter space open for the interpretation of black hole near-horizon physics, as these regimes remain poorly constrained and subject to significant uncertainties.\\\\ The stability of Frolov black holes has previously been investigated by computing QNMs under scalar perturbations \cite{Song:2024kkx}. In \cite{Lopez:2018aec}, the authors investigated the properties of QNMs for a probe massless scalar field in the background of Frolov BH in the eikonal limit. In this work, we aim to extend this analysis to electromagnetic and Dirac field perturbations, examining both the fundamental quasinormal mode and the first overtone during the ringdown phase. In addition, we compute the associated greybody factors, which quantify the transmission probability of Hawking radiation through the black hole’s effective potential barrier and directly influence its emission spectrum. Such a comprehensive analysis provides deeper insights into the near-horizon geometry and potential quantum gravity effects encoded in the QNM spectrum. More importantly, the possibility of detecting quasinormal ringing signatures from black holes has been proposed in the context of future space-based gravitational-wave observatories, such as LISA, which will be capable of probing these frequencies with high precision as discussed in \cite{LISA:2022kgy, Barausse:2020rsu}. Consequently, our results may offer a window into testing quantum gravity corrections to black hole spacetimes through gravitational wave observations.\\\\ The paper has been organized in the following manner: In Section (\ref{sec:GR1}), we describe the Frolov Black Hole metric, in Section (\ref{sec:GR2}), we work out the perturbation equations for the electromagnetic and Dirac field cases, in Section (\ref{sec:GR3}) we describe the WKB method and compute the QNM frequencies for the given perturbations in Tables (\ref{tab:QNM_Frolov1}) to (\ref{tab:QNM_Frolov8}). In Section (\ref{sec:GR4}) we compute the grey-body factors associated with the Frolov BH and in Section (\ref{sec:GR6}) we work out the Unruh-Verlinde temperature and compare how quantum corrections modify the standard results.

\section{The Frolov Black Hole Metric}
\label{sec:GR1}


A Frolov black hole can be viewed as a charged extension of the Hayward BH, 
first proposed in Ref.~\cite{Frolov:2016pav}.  
Its spacetime geometry is given by
\begin{equation}
    ds^{2} = -f(r)\,dt^{2} + \frac{dr^{2}}{f(r)} + r^{2} d\theta^{2} + r^{2}\sin^{2}\theta\, d\phi^{2},
    \label{eq:metric}
\end{equation}
where
\begin{equation}
    f(r) = 1 - \frac{(2 M r - q^{2}) r^{2}}{r^{4} + (2 M r + q^{2}) \alpha_{0}^{2}},
    \label{eq:f_r}
\end{equation}
with $M$ the black-hole mass.  
The central region of a Frolov BH behaves as if it were endowed with an effective cosmological constant 
$\Lambda = 3/\alpha_{0}^{2}$, where $\alpha_{0}$ represents the Hubble length.  
This Hubble length functions as a form of ``universal hair'' and is subject to the bound~\cite{Hayward:2005gi}
\begin{equation}
    \alpha_{0} \le \sqrt{\frac{16}{27} M}.
    \label{eq:bound}
\end{equation}
Meeting this condition implies that quantum-gravity effects become significant.  
For convenience we set $M=1$ in what follows, with no loss of generality.

The charge parameter $q$ introduces a specific black-hole ``hair'' and satisfies $0 \le q \le 1$.  
When $q = 0$, the Frolov BH reduces to the Hayward solution, whereas setting $\alpha_{0} = 0$ 
recovers the Reissner--Nordström (RN) geometry.  
If both $q = 0$ and $\alpha_{0} = 0$, the metric further simplifies to the Schwarzschild solution.

\begin{figure}[h!]
    \centering
    \includegraphics[width=0.48\textwidth]{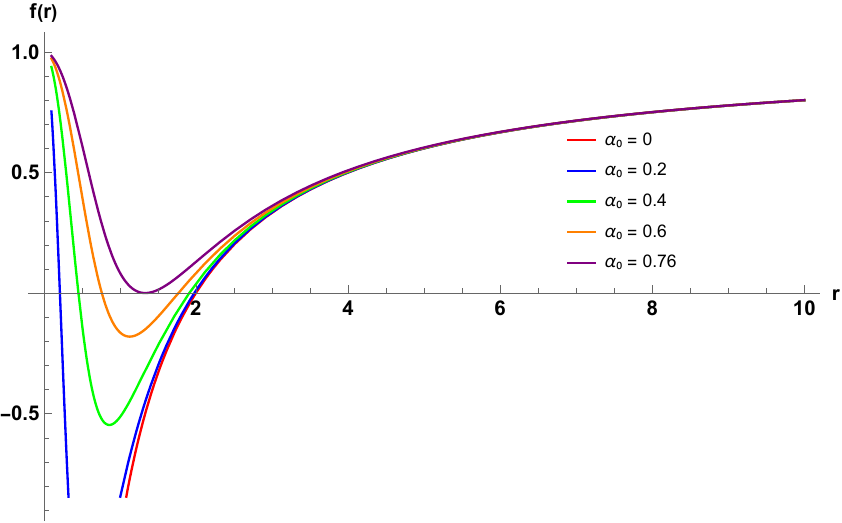}
    \caption{The metric function $f(r)$ as a function of $r$ for various values of $\alpha_0$ and with a fixed $q =0$
    Increasing \(\alpha_0\) shifts the curve upward and modifies the horizon structure.}
    \label{fr_plot1}
\end{figure}

\begin{figure}[h!]
    \centering
    \includegraphics[width=0.48\textwidth]{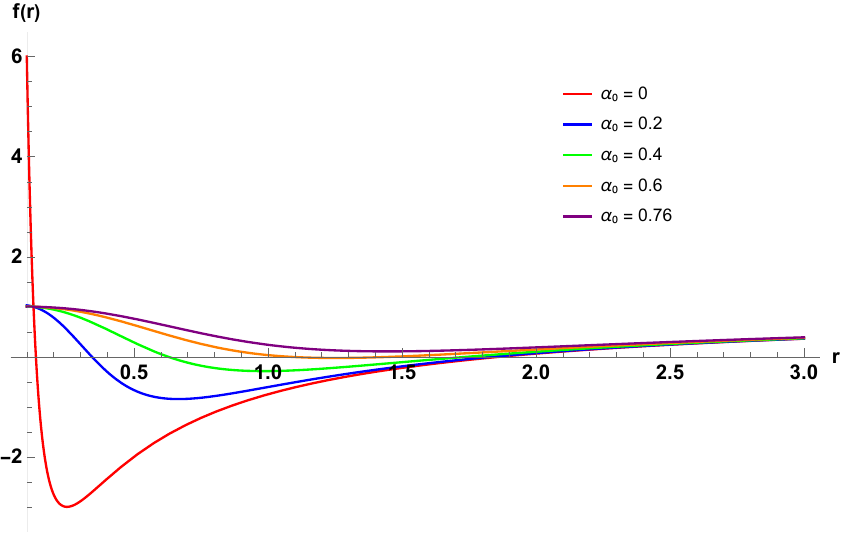}
    \caption{The metric function $f(r)$ as a function of $r$ for various values of $\alpha_0$ and with a fixed $q =0.5$
    Increasing \(\alpha_0\) shifts the curve upward and modifies the horizon structure.}
    \label{fr_plot2}
\end{figure}

\begin{figure}[h!]
    \centering
    \includegraphics[width=0.48\textwidth]{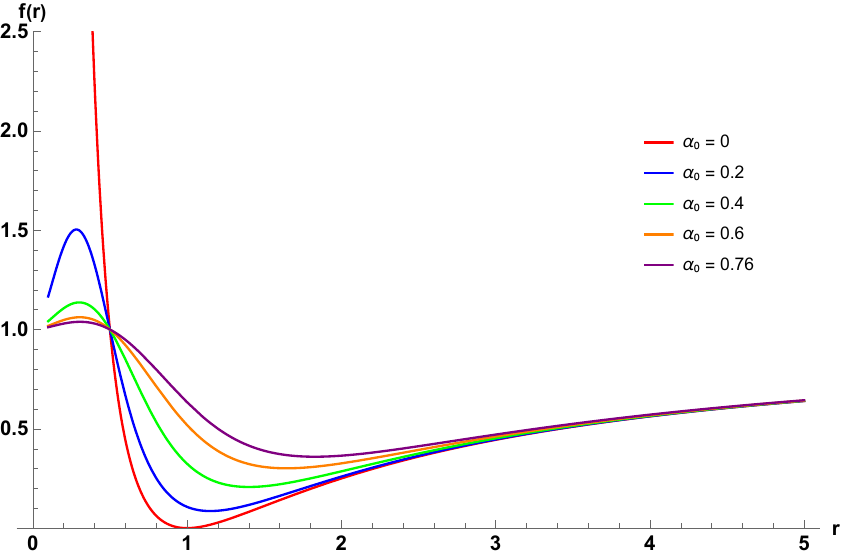}
    \caption{The metric function $f(r)$ as a function of $r$ for various values of $\alpha_0$ and with a fixed $q = 1$
    Increasing \(\alpha_0\) shifts the curve upward and modifies the horizon structure.}
    \label{fr_plot3}
\end{figure}

Figures~\ref{fr_plot1}, \ref{fr_plot2}, \ref{fr_plot3} illustrate the metric function $f(r)$ for various choices of $q$ and $\alpha_{0}$.  
As an example, consider $q=0$, which corresponds to the Hayward BH.  
Starting from the Schwarzschild case and increasing $\alpha_{0}$, the Hayward geometry develops a pair of horizons.  
Further increasing $\alpha_{0}$ up to the upper limit given by Eq.~\eqref{eq:bound} leads to the formation of a 
black hole with double horizons.

\section{Perturbation equations}
\label{sec:GR2}

The parameter $q$ appearing in the metric need not be interpreted uniquely as an electric charge; it may instead be associated with an effective or topological charge arising from non-electromagnetic sources, which justifies treating the electromagnetic field considered here as a test field. Therefore, we present below the field equations for the test electromagnetic and Dirac field perturbations in the given curved background.

The general covariant equation for an electromagnetic field is given by
\begin{equation}
    \frac{1}{\sqrt{-g}} \partial_{\mu} 
    \left( F_{\rho\sigma} g^{\rho\nu} g^{\sigma\mu} \sqrt{-g} \right) = 0,
\end{equation}
where $F_{\rho\sigma} = \partial_{\rho} A_{\sigma} - \partial_{\sigma} A_{\rho}$ 
and $A_{\mu}$ is the electromagnetic four-potential. It is worth noting that electromagnetic perturbations are treated in the test-field approximation, with backreaction effects and coupled gravito-electromagnetic perturbations neglected.

The vector potential $A_{\mu}$ can be expanded in terms of four-dimensional vector spherical harmonics 
(see Ref.~\cite{DeWitt:1973uma}) as

\begin{widetext}

\begin{equation}
    A_{\mu}(t,r,\theta,\phi) = 
    \sum_{\ell, m}
    \left[
    \begin{pmatrix}
        0 \\
        0 \\
        a_{\ell m}(t,r) \sin^{-1}\theta \,\partial_{\phi} Y_{\ell m} \\
        - a_{\ell m}(t,r) \sin\theta \,\partial_{\theta} Y_{\ell m}
    \end{pmatrix}
    +
    \begin{pmatrix}
        f_{\ell m}(t,r) Y_{\ell m} \\
        h_{\ell m}(t,r) Y_{\ell m} \\
        k_{\ell m}(t,r) \partial_{\theta} Y_{\ell m} \\
        k_{\ell m}(t,r) \partial_{\phi} Y_{\ell m}
    \end{pmatrix}
    \right],
    \label{eq:vector_potential}
\end{equation}
\end{widetext}
where $\ell$ is the angular quantum number and $m$ is the azimuthal number. 
The first column has parity $(-1)^{\ell+1}$ and the second has parity $(-1)^{\ell}$.

After separation of variables, the perturbation equations reduce to a Schr\"odinger-like wave equation 
of the form
\begin{equation}
    \frac{d^{2} \Psi_{\mathrm{EM}}}{dr_{*}^{2}} + 
    \left[\omega^{2} - V_{\mathrm{EM}}(r_*)\right] \Psi_{\mathrm{EM}} = 0,
\end{equation}
where $r_{*}$ is the tortoise coordinate, which is defined as:
\begin{equation}
    r_{*} = \int \frac{dr}{f(r)},
    \label{tortoise}
\end{equation}

The effective potential for electromagnetic perturbations is given by
\begin{equation}
    V_{\mathrm{EM}}(r) = f(r) \frac{\ell (\ell+1)}{r^{2}}.
\end{equation}

Similarly, we work out the field equations for the Dirac perturbations. 
For a general curved background spacetime, the massless Dirac equation reads
\begin{equation}
    \gamma^{a} e^{\mu}_{a} \left( \partial_{\mu} + \Gamma_{\mu} \right) \Psi = 0,
    \label{eq:dirac_eq}
\end{equation}
where $\gamma^{a}$ are the Dirac matrices, $e^{\mu}_{a}$ is the inverse of the tetrad $e^{a}_{\mu}$ 
with $g_{\mu \nu} = \eta_{ab} e^{a}_{\mu} e^{b}_{\nu}$, and $\eta_{ab}$ is the Minkowski metric. 
The spin connections $\Gamma_{\mu}$ are given by
\begin{equation}
    \Gamma_{\mu} = \frac{1}{8} \left[ \gamma^{a}, \gamma^{b} \right] e^{\nu}_{a} e_{b \nu ; \mu},
    \qquad 
    e_{b \nu ; \mu} = \partial_{\mu} e_{b \nu} - \Gamma^{\alpha}_{\mu \nu} e_{b \alpha}
\end{equation}

To separate the Dirac equation, we choose the tetrad
\begin{equation}
    e^{a}_{\mu} = \mathrm{diag}\!\left( \sqrt{f}, \frac{1}{\sqrt{f}}, r, r \sin\theta \right).
\end{equation}
Substituting this tetrad into Eq.~\eqref{eq:dirac_eq}, the Dirac equation becomes

\begin{align}
 \frac{\gamma^{0}}{\sqrt{f}}\, \frac{\partial \psi}{\partial t}
 &+ \sqrt{f}\,\gamma^{1} \left( \frac{\partial}{\partial r} 
    + \frac{1}{r} + \frac{1}{4f}\frac{df}{dr} \right) \psi  \notag \\
 &+ \frac{\gamma^{2}}{r}\left( \frac{\partial}{\partial \theta}
     + \frac{1}{2} \cot\theta \right)\psi
 + \frac{\gamma^{3}}{r \sin\theta}\, \frac{\partial \psi}{\partial \varphi} 
 = 0
\end{align}

Defining the rescaled perturbation $\psi = f^{-1/4}\phi$, the equation becomes
\begin{align}
 \frac{\gamma^{0}}{\sqrt{f}}\, \frac{\partial \phi}{\partial t}
 &+ \sqrt{f}\,\gamma^{1} \left( \frac{\partial}{\partial r} + \frac{1}{r} \right)\phi \notag \\
 &+ \frac{\gamma^{2}}{r}\left( \frac{\partial}{\partial \theta}
     + \frac{1}{2} \cot\theta \right)\phi
 + \frac{\gamma^{3}}{r \sin\theta}\, \frac{\partial \phi}{\partial \varphi}
 = 0
\end{align}

The Pauli matrices $\sigma_{i}$ are defined as
\begin{equation}
    \sigma_{1} = 
    \begin{pmatrix}
        0 & 1 \\ 1 & 0
    \end{pmatrix}, \quad
    \sigma_{2} = 
    \begin{pmatrix}
        0 & -i \\ i & 0
    \end{pmatrix}, \quad
    \sigma_{3} = 
    \begin{pmatrix}
        1 & 0 \\ 0 & -1
    \end{pmatrix}.
\end{equation}

Invoking the tortoise coordinate from eqn (\ref{tortoise}) and the ansatz for the Dirac spinor
\begin{equation}
    \phi(t,r,\theta,\varphi) = 
    \begin{pmatrix}
        \displaystyle i \frac{G^{(\pm)}(r)}{r} \chi^{\pm}_{jm}(\theta,\varphi) \\[6pt]
        \displaystyle \frac{F^{(\pm)}(r)}{r} \chi^{\mp}_{jm}(\theta,\varphi)
    \end{pmatrix}
    e^{-i \omega t},
\end{equation}
where $\chi^{\pm}_{jm}$ are spinor spherical harmonics given by
\begin{equation}
    \chi^{+}_{jm} =
    \begin{pmatrix}
        \sqrt{\frac{j+m}{2j}}\,Y_{\ell}^{m-1/2} \\[6pt]
        \sqrt{\frac{j-m}{2j}}\,Y_{\ell}^{m+1/2}
    \end{pmatrix}, 
    \qquad j = \ell + \frac{1}{2},
\end{equation}
and
\begin{equation}
    \chi^{-}_{jm} =
    \begin{pmatrix}
        \sqrt{\frac{j+1-m}{2j+2}}\,Y_{\ell}^{m-1/2} \\[6pt]
        -\sqrt{\frac{j+1+m}{2j+2}}\,Y_{\ell}^{m+1/2}
    \end{pmatrix}, 
    \qquad j = \ell - \frac{1}{2}.
\end{equation}
Here $Y_{\ell}^{m \pm 1/2}(\theta,\varphi)$ are the usual spin-weighted spherical harmonics. We proceed with the separation of variables by using a property of the spinor spherical harmonics. The angular operator appearing in eqn.(13) acts on $\chi^{\pm}_{jm}$ as an eigenvalue operator. In particular, the spinor harmonics satisfy the following relation, which allows the angular dependence of the Dirac equation to be separated:


\begin{equation}
\begin{split}
&\left(
-i\,
\scalebox{0.92}{$
\begin{pmatrix}
\displaystyle \frac{\partial}{\partial\theta} + \tfrac{1}{2}\cot\theta &
\displaystyle \frac{1}{\sin\theta}\frac{\partial}{\partial\varphi} \\[8pt]
\displaystyle -\frac{1}{\sin\theta}\frac{\partial}{\partial\varphi} &
\displaystyle -\!\left(\frac{\partial}{\partial\theta} + \tfrac{1}{2}\cot\theta\right)
\end{pmatrix}
$}
\right)
\begin{pmatrix}
\chi^{\pm}_{jm} \\[4pt] \chi^{\mp}_{jm}
\end{pmatrix}
\\[6pt]
&\qquad
= i\,
\scalebox{0.92}{$
\begin{pmatrix}
k_{\pm} & 0\\[4pt] 0 & k_{\pm}
\end{pmatrix}
$}
\begin{pmatrix}
\chi^{\pm}_{jm} \\[4pt] \chi^{\mp}_{jm}
\end{pmatrix}.
\end{split}
\end{equation}

\begin{equation}
\begin{split}
&\Bigg[
\begin{pmatrix}
0 & -\omega \\[4pt]
\omega & 0
\end{pmatrix}
\begin{pmatrix}
F^{\pm} \\[4pt] G^{\pm}
\end{pmatrix}
- \frac{\partial}{\partial r_{*}}
\begin{pmatrix}
F^{\pm} \\[4pt] G^{\pm}
\end{pmatrix} \\[6pt]
&\qquad
+ \sqrt{f}\,
\scalebox{0.92}{$
\begin{pmatrix}
\frac{k_{\pm}}{r} & 0 \\[6pt]
0 & -\frac{k_{\pm}}{r}
\end{pmatrix}$}
\begin{pmatrix}
F^{\pm} \\[4pt] G^{\pm}
\end{pmatrix}
\Bigg] = 0.
\end{split}
\end{equation}

The cases $(+)$ and $(-)$ in the functions can be put together after some matching, and the equation can be decoupled as
\begin{align}
\frac{d^{2}F}{dr_{*}^{2}} + \left(\omega^{2} - V_{1}\right) F &= 0, \label{eq:diracF} \\
\frac{d^{2}G}{dr_{*}^{2}} + \left(\omega^{2} - V_{2}\right) G &= 0. \label{eq:diracG}
\end{align}

After separation of variables, the Dirac equation reduces to a Schr\"odinger-like wave equation of the form
\begin{equation}
    \frac{d^{2} \Psi_{s}}{dr_{*}^{2}} + 
    \left[ \omega^{2} - V_{s}(r) \right] \Psi_{s} = 0,
    \label{eq:dirac_wave_eq}
\end{equation}
where $r_{*}$ is the tortoise coordinate and $V_{s}(r)$ is the effective potential for the Dirac perturbation. 
In our case, the potential takes the form
\begin{equation}
    V_{\pm 1/2}(r) = 
    \sqrt{f}\,\frac{|k|}{r^{2}} 
    \left( |k| \sqrt{f} \pm \frac{r}{2}\frac{df}{dr} \mp f \right),
    \label{eq:dirac_potential}
\end{equation}
where $|k| = 1, 2, 3, \dots$ is the total angular momentum quantum number. 
As shown in Ref.~\cite{Li:2013fka}, for a generic spherically symmetric spacetime, 
the two potentials $V_{+1/2}(r)$ and $V_{-1/2}(r)$ are isospectral, 
i.e., they yield identical quasinormal mode spectra. 

In our analysis, we therefore work exclusively with $V_{+1/2}(r)$, 
which is also more convenient for semi-analytic methods such as the WKB approximation.

\begin{figure}[h!]
    \centering
    \includegraphics[width=0.48\textwidth]{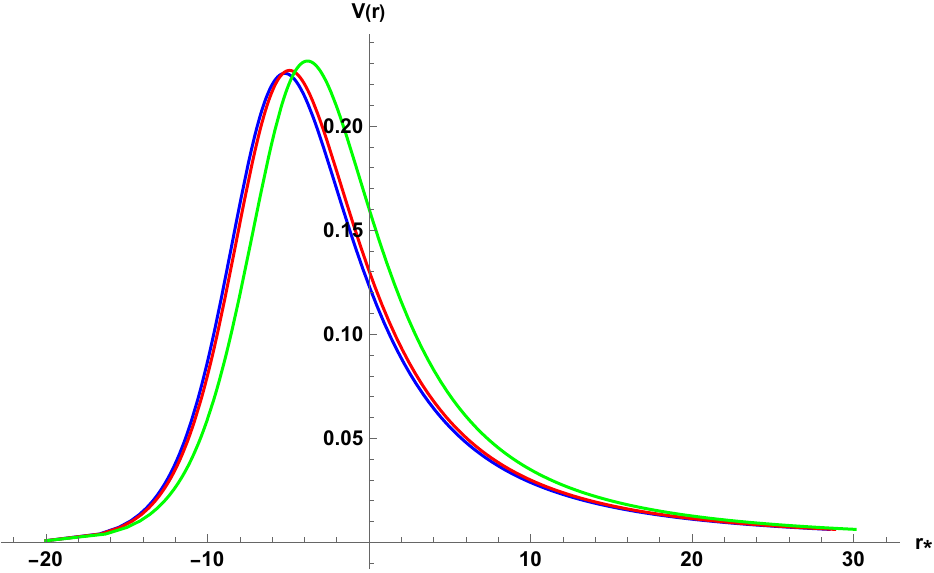}
    \caption{The variation of $V(r)$ with the tortoise coordinate $r_\ast$ for varying values of $\alpha_0$ = 0 (red), 0.2 (blue), 0.4 (green), 0.76 (purple) taking $q=0.2$, $l = 2$, $M=1$ for the massless electromagnetic perturbations.}
\label{Pot1}
\end{figure}

\begin{figure}[h!]
    \centering
    \includegraphics[width=0.6\textwidth]{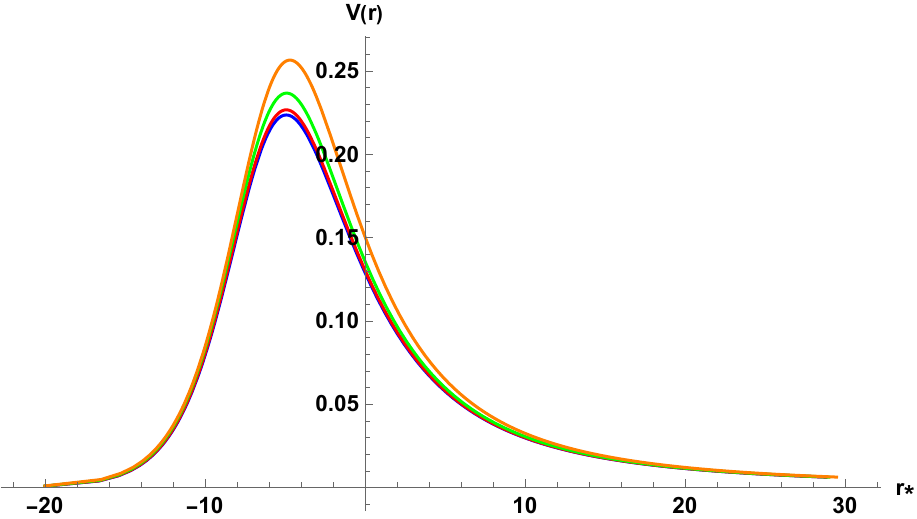}
    \caption{The variation of $V(r)$ with the tortoise coordinate $r_\ast$ for different values of $q$ = 0 (red), 0.3 (blue), 0.6 (green) with fixed angular momentum number $l = 2$, and $\alpha_0 = 0.2$ for the massless electromagnetic perturbations, taking $M = 1$.}

    \label{Pot2}
\end{figure}

\begin{figure}[h!]
    \centering
    \includegraphics[width=0.48\textwidth]{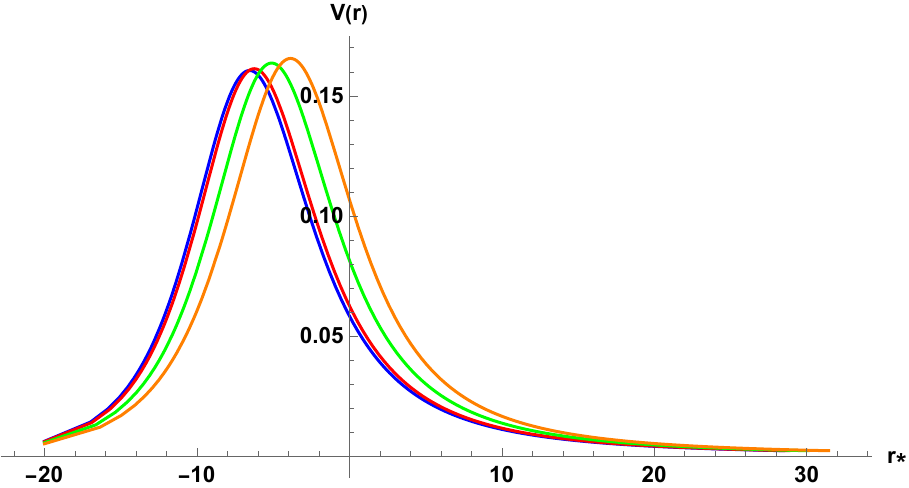}
    \caption{The variation of $V(r)$ with the tortoise coordinate $r_\ast$ for the Dirac field perturbations with $|k| = 2$ and varying  $\alpha_0$  = 0 (red), 0.2 (blue), 0.4 (green), 0.5 (orange) and $M = 1$.}

    \label{Pot3}
\end{figure}

\begin{figure}[h!]
    \centering
    \includegraphics[width=0.5\textwidth]{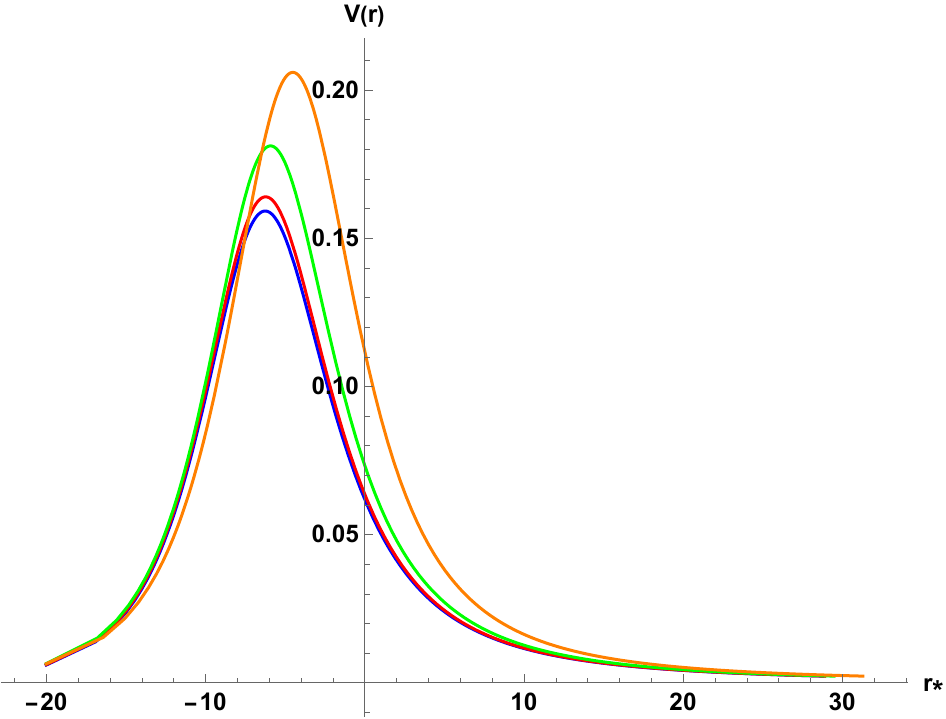}
   \caption{The variation of $V(r)$ with the tortoise coordinate $r_\ast$ for the Dirac field perturbations with $|k| = 2$, considering different values of $q$ = 0 (red), 0.3 (blue), 0.6 (green), 0.8 (orange), and $\alpha = 0.2$, $M = 1$.}

    \label{Pot4}
\end{figure}

\vspace{4mm}

\section{WKB method and QNM frequencies}
\label{sec:GR3}

The computation of QNM frequencies reduces to solving an eigenvalue problem for the perturbation equation, subject to well-defined boundary conditions. These conditions are chosen to represent a physical response of the spacetime to a transient disturbance. We impose
\begin{equation}
\psi(r_*) \sim e^{\pm i\omega r_*}, \quad r_* \to \pm\infty, 
\label{BC1}
\end{equation}
where $r_*$ is the tortoise coordinate. The choice $e^{+i\omega r_*}$ at $r_* \to +\infty$ corresponds to a purely outgoing wave at spatial infinity, while $e^{-i\omega r_*}$ at $r_* \to -\infty$ represents a purely ingoing wave at the event horizon. Physically, these conditions encode the idea that no radiation can emerge from behind the horizon and that all perturbations detected far away are purely outgoing. They describe the characteristic ``ringdown'' stage of a black hole's response, which is the phase that follows after any external source of perturbation has ceased \cite{Kokkotas:1999bd}. These boundary conditions select a discrete set of complex frequencies $\omega = \omega_R + i\omega_I$, where $\omega_R$ represents the oscillation frequency and $\omega_I < 0$ encodes the damping rate. 

The evolution of the test field perturbations in the Frolov BH background is governed by the master equation,
\begin{equation}
\frac{d^2\psi}{dx^2} + Q(x)\psi = 0,
\label{DE1}
\end{equation} 
where $x$ is the tortoise coordinate $r_*$, and $Q(x)$ is defined as
\begin{equation}
Q(x) = \omega^2 - V_{\text{eff}}(r_*). 
\end{equation}
The functional form of $Q(x)$ influences the structure of the effective potential in the master equation; the actual decay or growth of perturbations is determined only after specifying the appropriate initial data and boundary conditions which are discussed above.

To compute the QNM frequencies and analyze the stability of the Frolov black hole under given test field perturbations, 
we use the WKB approximation. This semi-analytic technique is particularly well-suited for potentials 
with a barrier-like structure, as is the case here, where the Regge--Wheeler (RW) potential 
governs the evolution of perturbations (see Fig.~\ref{Pot1}, \ref{Pot2}, \ref{Pot3}, and \ref{Pot4}). 

The Schr\"odinger-like equation in curved spacetime was first solved using the WKB method 
by Schutz and Will in 1985~\cite{Schutz_1985}. 
Iyer and Will subsequently extended the formalism to third order in 1987~\cite{Iyer_1987}, 
significantly improving its precision. 
Later, Konoplya extended the method up to sixth order~\cite{Konoplya:2003ii}.
While higher-order WKB schemes generally improve accuracy, the WKB expansion itself is asymptotic and may fail to converge at higher orders. To mitigate this limitation, one typically employs Padé approximants, following Matyjasek and Opala ~\cite{Matyjasek:2017psv}, which enhance the convergence properties of the WKB series.
It is worth noting that the results obtained at very high orders, such as the thirteenth, sometimes deviate 
significantly from those at lower orders due to error amplification and lack of convergence~\cite{Hatsuda:2019eoj}.

In this work, we focus primarily on the sixth and eighth-order WKB approximations, which strikes a good balance between accuracy and computational simplicity. The general WKB expression for the frequencies can be expressed as an expansion around the eikonal limit, as follows
\begin{equation}
\begin{aligned}
\omega^{2} =\;&
V_{0}
+ A_{2}(K^{2})
+ A_{4}(K^{2})
+ A_{6}(K^{2})
+ \cdots \\
& \hspace{-8mm} - i K \sqrt{-2 V_{0}''}
\Bigl(
1
+ A_{3}(K^{2})
+ A_{5}(K^{2})
+ A_{7}(K^{2})
+ \cdots
\Bigr)
\label{wkbformula}
\end{aligned}
\end{equation}
where the matching conditions for the QNM imply a quantization condition,
\begin{equation}
K = n + \frac{1}{2}, \qquad n = 0,1,2,\ldots 
\end{equation}
with $n$ being the overtone number. Here, $V_0$ denotes the value of the effective potential at its maximum, $V_0''$ represents the second derivative of the potential at this point with respect to the tortoise coordinate, and $A_i$ for $i=2,3,4,\ldots$ signify the $i$th-order WKB correction terms.

To improve the accuracy of higher-order WKB calculations, we follow the Padé averaging procedure presented in Ref. \cite{Konoplya:2019hlu}. Within this approach, one introduces a WKB polynomial \(P_k(\varepsilon)\) by inserting a formal order-counting parameter \(\varepsilon\) into the right-hand side of the WKB expansion, such that
\begin{equation}
\begin{aligned}
P_k(\varepsilon) =\;&
V_0
+ A_2(K^2)\,\varepsilon^2
+ A_4(K^2)\,\varepsilon^4
+ A_6(K^2)\,\varepsilon^6
+ \cdots \\
& \hspace{-7mm}- iK \sqrt{-2V_0''}
\left(
\varepsilon
+ A_3(K^2)\,\varepsilon^3
+ A_5(K^2)\,\varepsilon^5
+ \cdots
\right),
\end{aligned}
\end{equation}
where the polynomial order \(k\) coincides with the order of the WKB approximation, and a formal bookkeeping parameter \(\varepsilon\) is introduced to organize the WKB expansion as an asymptotic series and to keep track of successive orders in the WKB approximation. The squared quasinormal frequency is obtained by setting \(\varepsilon = 1\),
\begin{equation}
\omega^2 = P_k(1).
\end{equation}

We consider a family of the Padé approximants \(P^{\tilde n}_{\tilde m}(\varepsilon)\) for the polynomial $P_k(\varepsilon)$ near $\varepsilon =0$ with \(\tilde n+\tilde m=k\),
\begin{equation}
P^{\tilde n}_{\tilde m}(\varepsilon)=\frac{Q_0+Q_1\varepsilon+\cdots+Q_{\tilde n}\varepsilon^{\tilde n}}
{R_0+R_1\varepsilon+\cdots+R_{\tilde m}\varepsilon^{\tilde m}},
\end{equation}
so that \(P^{\tilde n}_{\tilde m}(\varepsilon)-P_k(\varepsilon)=\mathcal{O}(\varepsilon^{k+1})\).
The squared frequency is approximated by evaluating the rational function at \(\varepsilon=1\),
\[
\omega^2 \approx P^{\tilde n}_{\tilde m}(1).
\]
Different choices of \((\tilde n,\tilde m)\) therefore give a family of approximations \(\{\omega_{\tilde n/\tilde m}\}\) at the same WKB order.

To form a reasonable estimate for order \(k\) we use the (ad-hoc) averaging recipe:
\begin{enumerate}
  \setlength{\itemsep}{2pt}
  \setlength{\parsep}{0pt}
  \setlength{\topsep}{2pt}
  \setlength{\partopsep}{0pt}
  
   \item Compute the set of Pad\'e results $\omega_{\tilde n/\tilde m}$ with $\tilde n + \tilde m = k$.
  
  \item Define the central value $\omega_k^{(c)}$ (roughly $\tilde n \approx \tilde m$) and the mean of central values $\omega_k^{(m)}$.
  
  \item Identify the $r=\lfloor (k+1)/3 \rfloor$ pairs of closest values and compute the averages $\omega_k^{(1)}, \omega_k^{(2)}, \ldots$.
  
  \item Adopt as the order-$k$ estimate $\omega_k$ the mean of those $r$ closest values and estimate the error by the sample standard deviation $S_k$ of the same set.
\end{enumerate}

This procedure typically improves the accuracy compared to the non-Pad\'e WKB result and provides an internal error estimate via. For details, explicit formulas, examples and caveats (including cases where some Pad\'e approximants produce outliers and when extra turning points spoil the WKB applicability) see \cite{Konoplya:2019hlu}.

However, the presence of extremely large higher-order terms can lead to numerical instabilities, 
as observed for the twelfth-order approximation. 
Our analysis shows that for the electromagnetic and the Dirac perturbation case, we have employed the sixth-order WKB approximation. To ensure numerical consistency, we have also cross-checked our results with higher-order values as well as the Padé Averaged results \cite{Matyjasek:2019eeu}, specifically the eighth-order for the electromagnetic, and for the Dirac case. The outcomes from different orders are found to be in good agreement across all the considered values of the parameters in the metric.
In this work, we implement the WKB formalism using the publicly available numerical 
package developed by R.A. Konoplya~\cite{Konoplya202X}.

\begin{widetext}

\begin{table}[htbp]
\centering
\caption{Quasinormal mode frequencies $\omega$ for the massless electromagnetic field perturbations for the fundamental mode ($n=0$), obtained using the 6th and 8th order WKB methods with Padé Averaging for various values of $q$ and $\alpha_0$. We set $M=1$ and $l =2$.}
\label{tab:QNM_Frolov1}
\renewcommand{\arraystretch}{1.3} 
\setlength{\tabcolsep}{9pt}       
\begin{tabular}{c c c c c c}
\hline\hline
$q$ & $\alpha_0$ & 
\multicolumn{2}{c}{6th order WKB (Padé), $(n=0)$} & 
\multicolumn{2}{c}{8th order WKB (Padé), $(n=0)$} \\
\cline{1-6}
 &  & $Re(\omega)$ & $Im(\omega)$ & $Re(\omega)$ & $Im(\omega)$ \\
\hline
0.0  &  0.0    &    0.457595           &   $- 0.0950048i$            &    0.457594          & $- 0.0950038i$             \\
   0.2  &   0.0   &   0.460828           &  $-0.0952257i$            &   0.460828            &     $-0.0952258i$         \\
 0.2    &   0.1   &     0.461245         &    $-0.0950674i$          &    0.461245           & $-0.0950674i$             \\
0.4  &  0.2    &     0.473106         &     $-0.0950677i$         &     0.473106         &   $-0.0950678i$           \\
  0.6   &   0.3   &      0.496867        &  $-0.0939648i$            &      0.496867         &      $-0.093965i$        \\
 0.8    &   0.4   &    0.54452           &  $-0.084335i$            &       0.544521       & $-0.0843349i$             \\

\hline\hline
\end{tabular}
\end{table}

\begin{table}[htbp]
\centering
\caption{Quasinormal mode frequencies $\omega$ for the massless electromagnetic field perturbations for the first overtone ($n=1$), obtained using the 6th and 8th order WKB methods with Padé Averaging for various values of $q$ and $\alpha_0$. We set $M=1$ and $l =2$.}
\label{tab:QNM_Frolov2}
\renewcommand{\arraystretch}{1.3} 
\setlength{\tabcolsep}{9pt}       
\begin{tabular}{c c c c c c}
\hline\hline
$q$ & $\alpha_0$ & 
\multicolumn{2}{c}{6th order WKB (Padé), $(n=1)$} & 
\multicolumn{2}{c}{8th order WKB (Padé), $(n=1)$} \\
\cline{1-6}
 &  & $Re(\omega)$ & $Im(\omega)$ & $Re(\omega)$ & $Im(\omega)$ \\
\hline
0.0  &  0.0    &    0.436533           &   $-0.290727i$            &    0.436522         & $-0.290724i$             \\
   0.2  &   0.0   &   0.439932           &  $-0.291343i$            &   0.439932             &     $-0.291343i$         \\
 0.2    &   0.1   &     0.440478         &    $-0.290827i$          &    0.440477           & $-0.290827i$             \\
0.4  &  0.2    &     0.453322          &     $-0.290489i$         &     0.453324      &   $-0.290491i$           \\
  0.6   &   0.3   &     0.478945       &  $-0.28627i$            &      0.478945          &      $-0.28627i$        \\
 0.8    &   0.4   &    0.521604            &  $-0.256562i$            &       0.521576       & $-0.256573i$             \\

\hline\hline
\end{tabular}
\end{table}

\begin{table}[htbp]
\centering
\caption{Quasinormal mode frequencies $\omega$ for the massless electromagnetic field perturbations for the fundamental mode ($n=0$), obtained using the 6th and 8th order WKB methods with Padé Averaging for various values of $q$ and $\alpha_0$. We set $M=1$ and $l =3$.}
\label{tab:QNM_Frolov3}
\renewcommand{\arraystretch}{1.3} 
\setlength{\tabcolsep}{9pt}       
\begin{tabular}{c c c c c c}
\hline\hline
$q$ & $\alpha_0$ & 
\multicolumn{2}{c}{6th order WKB (Padé), $(n=0)$} & 
\multicolumn{2}{c}{8th order WKB (Padé), $(n=0)$} \\
\cline{1-6}
 &  & $Re(\omega)$ & $Im(\omega)$ & $Re(\omega)$ & $Im(\omega)$ \\
\hline
0.0  &  0.0    &    0.656899            &   $-0.0956163i$            &      0.656899       & $-0.0956162i$            \\
   0.2  &   0.0   &    0.661439        &  $-0.0958322i$            &    0.661439            &   $-0.0958324i$       \\
 0.2    &   0.1   &     0.66199         &   $-0.0956758i$           &    0.66199           & $-0.0956758i$             \\
0.4  &  0.2    &     0.678516         &   $-0.0956705i$           &    0.678516       &    $-0.0956705i$          \\
  0.6   &   0.3   &  0.711544      &    $-0.0945891i$      &      0.711544           &  $-0.0945891i$           \\
 0.8    &   0.4   &    0.778514          &   $-0.0852404i$           &      0.778515        &      $-0.0852402i$        \\

\hline\hline
\end{tabular}
\end{table}

\begin{table}[htbp]
\centering
\caption{Quasinormal mode frequencies $\omega$ for the massless electromagnetic field perturbations for the first overtone ($n=1$), obtained using the 6th and 8th order WKB methods with Padé Averaging for various values of $q$ and $\alpha_0$. We set $M=1$ and $l =3$.}
\label{tab:QNM_Frolov4}
\renewcommand{\arraystretch}{1.3} 
\setlength{\tabcolsep}{9pt}       
\begin{tabular}{c c c c c c}
\hline\hline
$q$ & $\alpha_0$ & 
\multicolumn{2}{c}{6th order WKB (Padé), $(n=1)$} & 
\multicolumn{2}{c}{8th order WKB (Padé), $(n=1)$} \\
\cline{1-6}
 &  & $Re(\omega)$ & $Im(\omega)$ & $Re(\omega)$ & $Im(\omega)$ \\
\hline
0.0  &  0.0    &    0.641736           &   $-0.289731i$            &      0.641735       & $-0.28973i$            \\
   0.2  &   0.0   &    0.646396         &  $-0.290355i$            &   0.646396            &   $-0.290354i$       \\
 0.2    &   0.1   &     0.647034        &   $-0.289862i$           &    0.647034         & $-0.289862i$             \\
0.4  &  0.2    &     0.664249        &   $-0.289673i$           &    0.664249      &    $-0.289674i$          \\
  0.6   &   0.3   &  0.698578      &    $-0.285969i$      &     0.698578           &  $-0.285969i$           \\
 0.8    &   0.4   &    0.762154         &   $-0.257542i$           &      0.762149        &      $-0.257543i$        \\

\hline\hline
\end{tabular}
\end{table}

\begin{table}[htbp]
\centering
\caption{Quasinormal mode frequencies $\omega$ for the massless Dirac field perturbations for the fundamental mode ($n=0$), obtained using the 6th and 8th order WKB methods with Padé Averaging for various values of $q$ and $\alpha_0$. We set $M=1$ and $|k| =2$.}
\label{tab:QNM_Frolov5}
\renewcommand{\arraystretch}{1.3} 
\setlength{\tabcolsep}{9pt}       
\begin{tabular}{c c c c c c}
\hline\hline
$q$ & $\alpha_0$ & 
\multicolumn{2}{c}{6th order WKB (Padé), $(n=0)$} & 
\multicolumn{2}{c}{8th order WKB (Padé), $(n=0)$} \\
\cline{1-6}
 &  & $Re(\omega)$ & $Im(\omega)$ & $Re(\omega)$ & $Im(\omega)$ \\
\hline
0.0  &  0.0    &    0.380054           &   $-0.0963853i$            &      0.380051       & $-0.0963904i$            \\
   0.2  &   0.0   &    0.38268         &  $-0.0965934i$            &   0.382675            &   $-0.0966004i$       \\
 0.2    &   0.1   &     0.382981        &   $-0.0964316i$           &   0.382977          & $-0.0964365i$             \\
0.4  &  0.2    &     0.392473         &   $-0.096383i$           &    0.392468      &    $-0.0963857i$          \\
  0.6   &   0.3   &  0.411318    &    $-0.0951834i$      &     0.411291          &  $-0.095181i$           \\
 0.8    &   0.4   &    0.447367          &   $-0.0859194i$           &     0.447366         &      $-0.0859201i$        \\

\hline\hline
\end{tabular}
\end{table}

\begin{table}[htbp]
\centering
\caption{Quasinormal mode frequencies $\omega$ for the massless Dirac field perturbations for the first overtone ($n=1$), obtained using the 6th and 8th order WKB methods with Padé Averaging for various values of $q$ and $\alpha_0$. We set $M=1$ and $|k| =2$.}
\label{tab:QNM_Frolov6}
\renewcommand{\arraystretch}{1.3} 
\setlength{\tabcolsep}{9pt}       
\begin{tabular}{c c c c c c}
\hline\hline
$q$ & $\alpha_0$ & 
\multicolumn{2}{c}{6th order WKB (Padé), $(n=1)$} & 
\multicolumn{2}{c}{8th order WKB (Padé), $(n=1)$} \\
\cline{1-6}
 &  & $Re(\omega)$ & $Im(\omega)$ & $Re(\omega)$ & $Im(\omega)$ \\
\hline
0.0  &  0.0    &   0.355778            &   $-0.297269i$            &      0.355769        & $-0.297303i$            \\
   0.2  &   0.0   &    0.35858        &  $-0.297826i$            &   0.358571          &   $-0.297842i$       \\
 0.2    &   0.1   &     0.359029        &   $-0.297324i$           &   0.359024         & $-0.297324i$             \\
0.4  &  0.2    &     0.369693         &   $-0.29662i$           &    0.369687       &    $-0.296612i$          \\
  0.6   &   0.3   &  0.390698      &    $-0.29152i$      &     0.390255           &  $-0.29184i$           \\
 0.8    &   0.4   &    0.41957           &   $-0.262825i$           &    0.419348         &      $-0.263346i$        \\

\hline\hline
\end{tabular}
\end{table}

\begin{table}[htbp]
\centering
\caption{Quasinormal mode frequencies $\omega$ for the massless Dirac field perturbations for the fundamental mode ($n=0$), obtained using the 6th and 8th order WKB methods with Padé Averaging for various values of $q$ and $\alpha_0$. We set $M=1$ and $|k| =3$.}
\label{tab:QNM_Frolov7}
\renewcommand{\arraystretch}{1.3} 
\setlength{\tabcolsep}{9pt}       
\begin{tabular}{c c c c c c}
\hline\hline
$q$ & $\alpha_0$ & 
\multicolumn{2}{c}{6th order WKB (Padé), $(n=0)$} & 
\multicolumn{2}{c}{8th order WKB (Padé), $(n=0)$} \\
\cline{1-6}
 &  & $Re(\omega)$ & $Im(\omega)$ & $Re(\omega)$ & $Im(\omega)$ \\
\hline
0.0  &  0.0    &   0.574094            &   $-0.0963048i$            &     0.574094        & $-0.0963048i$            \\
   0.2  &   0.0   &    0.578015        &  $-0.0965155i$            &   0.578015          &   $-0.0965156i$       \\
 0.2    &   0.1   &     0.578467       &   $-0.0963577i$           &   0.578467       & $-0.0963577i$             \\
0.4  &  0.2    &     0.592637         &   $-0.0963316i$           &   0.592636       &    $-0.0963317i$          \\
  0.6   &   0.3   &  0.620832      &    $-0.0952188i$      &    0.620831           &  $-0.095218i$           \\
 0.8    &   0.4   &    0.676997            &   $-0.086072i$           &     0.676996         &      $-0.0860717i$        \\

\hline\hline
\end{tabular}
\end{table}

\begin{table}[htbp]
\centering
\caption{Quasinormal mode frequencies $\omega$ for the massless Dirac field perturbations for the first overtone ($n=1$), obtained using the 6th and 8th order WKB methods with Padé Averaging for various values of $q$ and $\alpha_0$. We set $M=1$ and $|k| =3$.}
\label{tab:QNM_Frolov8}
\renewcommand{\arraystretch}{1.3} 
\setlength{\tabcolsep}{9pt}       
\begin{tabular}{c c c c c c}
\hline\hline
$q$ & $\alpha_0$ & 
\multicolumn{2}{c}{6th order WKB (Padé), $(n=1)$} & 
\multicolumn{2}{c}{8th order WKB (Padé), $(n=1)$} \\
\cline{1-6}
 &  & $Re(\omega)$ & $Im(\omega)$ & $Re(\omega)$ & $Im(\omega)$ \\
\hline
0.0  &  0.0    &  0.557015           &   $-0.292715i$            &    0.557015        & $-0.292715i$            \\
   0.2  &   0.0   &    0.56107      &  $-0.293314i$            &   0.561071          &   $-0.293315i$       \\
 0.2    &   0.1   &     0.561614      &   $-0.292812i$           &   0.561614     & $-0.292813i$             \\
0.4  &  0.2    &     0.576534          &   $-0.292511i$           &  0.576534       &    $-0.29251i$          \\
  0.6   &   0.3   &  0.606124      &    $-0.288564i$      &    0.606113         &  $-0.288563i$           \\
 0.8    &   0.4   &    0.658189           &   $-0.260521i$           &    0.658168       &      $-0.260649i$        \\

\hline\hline
\end{tabular}
\end{table}

\end{widetext}

As we can see, Table (\ref{tab:QNM_Frolov1}) lists the fundamental-mode QNMs for the massless electromagnetic perturbation ($l=2$) computed with 6th- and 8th-order WKB with Padé averaging. Both WKB orders produce very similar values for $Re(\omega)$ and $Im(\omega)$, indicating excellent agreement between the two orders of the method for these parameters. As the charge $q$ or the BH parameter $\alpha_0$ is increased, the real part of the frequency shifts upward and the modes oscillate faster while the magnitude of the imaginary part decreases, i.e., modes begin to damp more slowly, and thus the damping times lengthen. The table therefore demonstrates that introducing charge and parameter stiffen the effective potential and prolong the decay time of the fundamental electromagnetic mode.

Similarly, Table (\ref{tab:QNM_Frolov2}) reports the first overtone ($n=1$) for the electromagnetic field at $l=2$. The overtone behaves qualitatively like the fundamental mode with respect to $q$ and $\alpha_0$ (increasing $Re(\omega)$ and decreasing the absolute damping), but it displays the expected overtone characteristics: relative to the fundamental the overtone is more strongly damped (more negative $Im(\omega)$) and shows slightly greater sensitivity to the WKB order. This increased sensitivity for $n=1$ is a known property of WKB approaches and is visible here as marginally larger differences between the 6th- and 8th-order Pad\'e results compared with the fundamental.

`In Table (\ref{tab:QNM_Frolov3}) we compute the fundamental electromagnetic QNMs for $l=3$. Compared with the $l=2$ fundamental (Table \ref{tab:QNM_Frolov1}), higher multipole index results in larger $Re(\omega)$ and a larger magnitude of $Im(\omega)$, consistent with a steeper and narrower effective potential at larger $l$. The dependence on $q$ and $\alpha_0$ follows the same pattern seen before: raising either parameter increases the oscillation frequency and reduces the damping rate. Agreement between the 6th- and 8th-order Pad\'e-averaged WKB remains close, supporting numerical stability for these higher multipoles in the parameter ranges shown.
The first overtone for $l=3$ in Table (\ref{tab:QNM_Frolov4}) again shows the overtone features, it is more strongly damped and typically has a somewhat different real and imaginary parts compared to $n=0$. Parameter trends with respect to $q$ and $\alpha_0$ remain consistent with the the analysis of the previous tables.

In Table (\ref{tab:QNM_Frolov5}) we list fundamental modes for massless Dirac perturbations with $|k|=2$. Qualitatively the Dirac spectrum mirrors the electromagnetic case: increasing $q$ or $\alpha_0$ pushes $Re(\omega)$ upward and decreases the damping rate $Im(\omega)$ in magnitude, i.e. oscillations become faster and longer-lived as the parameters increase. The 6th- and 8th-order WKB--Pad\'e results are in close agreement for the fundamental Dirac mode, indicating that the chosen WKB orders are sufficiently accurate for these modes and parameter ranges, and similarly, we have Table (\ref{tab:QNM_Frolov6}) where the first overtone of the Dirac field at $|k|=2$ shows that overtone pattern, i.e, the real part of the frequencies is slightly lower than $n=0$ case. The parameter dependence is consistent with the other tables; both $q$ and $\alpha_0$ reduce damping magnitude and increase oscillation frequency. Practically, this means Dirac overtones also become longer lived and somewhat higher in frequency with increased charge or the BH parameter.

Finally we have the Table (\ref{tab:QNM_Frolov7}) for Dirac perturbations at $|k|=3$ (fundamental) both the real and imaginary parts are larger in magnitude compared to $|k|=2$, consistent with the multipole hierarchy: larger angular index corresponds to a higher frequency, faster decaying response. The monotone trends with $q$ and $\alpha_0$ persist here as well. The 6th- and 8th-order WKB--Pad\'e columns remain in good agreement, reinforcing confidence in the computed fundamental-mode values at these multipoles.

The first overtone for $|k|=3$ in Table (\ref{tab:QNM_Frolov8}) exhibits the standard behavior relative to the fundamental: increased damping and a characteristic shift of the real part consistent with overtone structure. Here it is observed that increasing $q$ or $\alpha_0$ increases the oscillation frequency and reduces the damping rate in magnitude. Differences between 6th- and 8th-order results are slightly larger here than for fundamentals, but do not change the qualitative conclusions drawn from the table.\\

Across all Tables, the QNM spectra exhibit a clear and consistent pattern: the real part of the frequency $Re(\omega)$ increases with both the electric charge $q$ and the regularization parameter $\alpha_0$, while the imaginary part $Im(\omega)$ becomes less negative, indicating slower damping and longer-lived perturbations. Increasing the angular index ($l$ or $|k|$) raises both the oscillation frequency and the damping magnitude. The close agreement between the 6th- and 8th-order WKB--Padé results for the fundamental modes confirms the reliability and numerical stability of the method, while the expected mild sensitivity for higher overtones remains within acceptable limits. These results suggest that the inclusion of charge and parameter effects modifies the effective potential of the underlying BH geometry, producing higher frequencies and longer lived oscillations.

It is important to note that in the present analysis, we have restricted our computations to the modes satisfying $n < l$ (or $n < |k|$ for Dirac perturbations) and have not considered the special case $l = n = 0$, since the WKB method is applicable only when $l \geq n$. Furthermore, for higher overtone numbers ($n \geq l$), the WKB method progressively loses accuracy due to the breakdown of the eikonal approximation and the increasing influence of the inner potential well at smaller radii. In such regimes, more robust numerical approaches such as time-domain integration, continued-fraction (Leaver) methods, or the asymptotic iteration method are required to obtain reliable QNM frequencies. We leave this direction of research for future investigations.


\section{Grey-body Factors}
\label{sec:GR4}


Grey-body factors quantify the fraction of the initial Hawking radiation 
that successfully transmits through the effective potential barrier surrounding the black hole, rather than being reflected back toward the event horizon. To compute this quantity, we apply Hawking’s semiclassical formula, now modified with a grey-body factor, to estimate the radiation flux that reaches a distant observer.

As discussed in the literature, the contribution of gravitons to the 
total radiation flux is negligible, for instance, in the Schwarzschild 
case, gravitons account for less than \(2\%\) of the emitted 
radiation~\cite{Page:1976df}. Consequently, grey-body factors for test 
fields are sufficient to characterize the radiation spectrum. 
Moreover, these factors can be more influential than the Hawking 
temperature itself in determining the intensity of the observed 
flux~\cite{Konoplya:2019ppy}.

To compute the grey-body factors, we analyze the wave equation under 
scattering boundary conditions that allow for an incident wave from 
spatial infinity. Owing to the symmetry of the scattering problem, 
this is equivalent to studying a wave incident from the horizon. 
The appropriate boundary conditions for the field \(\Psi(r_*)\) are 
given by
\begin{equation}
\Psi(r_*) =
\begin{cases}
e^{-i \omega r_*} + R e^{i \omega r_*}, & r_* \to +\infty, \\[6pt]
T e^{-i \omega r_*}, & r_* \to -\infty,
\end{cases}
\label{BC2}
\end{equation}
where \(R\) and \(T\) are the reflection and transmission coefficients, 
respectively.

Because the effective potential has a single barrier-like peak and 
falls off monotonically toward both infinities, the WKB 
approximation~\cite{Konoplya:2003ii} can be reliably applied 
to compute $R$ and $T$. Since \(\omega^2\) is real, where $R$ and $T$ are the complex reflection and
transmission amplitudes, respectively. The normalization condition ensuring energy conservation is
\begin{equation}
|T|^{2} + |R|^{2} = 1.
\end{equation}
From the reflection coefficient, the transmission coefficient for 
each multipole number \(\ell\) can be obtained as
\begin{equation}
|\Gamma_\ell|^{2} = 1 - |R_\ell|^{2} = |T_\ell|^{2},
\end{equation}
where \(\Gamma_\ell\) represents the grey-body factor.

In our analysis, we employ the higher-order WKB 
formula~\cite{Konoplya:2019hlu, Matyjasek:2017psv} to obtain a precise estimate 
of the reflection and transmission coefficients. This approach, however, 
becomes unreliable at very low frequencies, where almost all of the wave 
is reflected and the contribution to the total energy flux is negligible. 
In this regime, we follow the standard approach of extrapolating the WKB 
result to small \(\omega\).

According to Refs. \cite{Iyer_1987, Konoplya:2003ii}, the reflection 
coefficient takes the form
\begin{equation}
|R| = \left( 1 + e^{-2 i \pi K} \right)^{-1/2}
\label{refcoefficient}
\end{equation}
where \(K\) is numerically determined from equation (\ref{wkbformula}).

Here, we would like to emphasize that in the QNM problem, the boundary conditions of purely ingoing waves at the event horizon and purely outgoing waves at spatial infinity lead to a discrete set of complex frequencies through a quantization condition. In contrast, the computation of grey-body factors corresponds to a scattering problem with real frequencies, where incident, reflected, and transmitted waves are allowed. In this case, the same WKB expansion evaluated near the maximum of the effective potential determines the parameter $K$ in equation (\ref{refcoefficient}).

\begin{figure}[h!]
    \centering
    \includegraphics[width=0.49\textwidth]{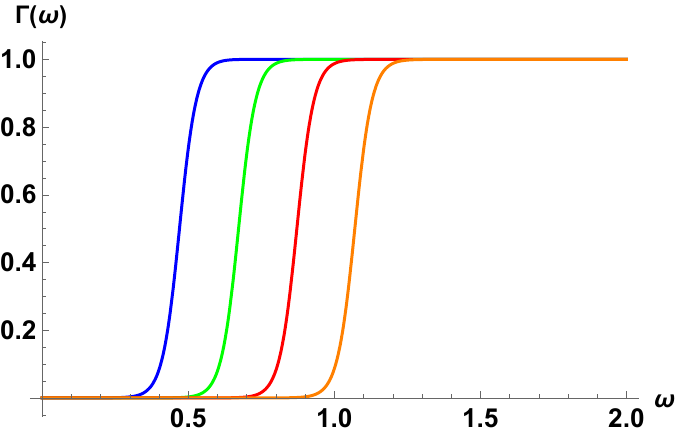}
    \caption{Grey-body factors of the massless electromagnetic field as a function of frequency for the Frolov black hole with $l = 2, 3, 4, 5$ (from left to right) and $\alpha_0 = 0.4$ and $q = 0.2$.}

    \label{GBF1}
\end{figure}

\begin{figure}[h!]
    \centering
    \includegraphics[width=0.49\textwidth]{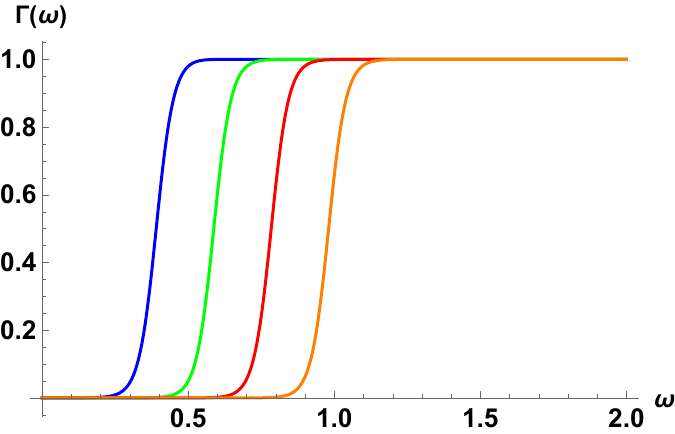}
   \caption{Grey-body factors of the Dirac field as a function of frequency for the Frolov black hole with $|k| = 2, 3, 4, 5$ (from left to right) and $\alpha = 0.4$ and $q = 0.2$.}

    \label{GBF2}
\end{figure}

We employ the $6{th}$ order approximation for calculating grey-body factors without using Pad\'e resummation, because it is not straightforward for grey-body factors, unlike the case of QNMs. The grey-body factors shown in Fig.~(\ref{GBF1}) and Fig.~(\ref{GBF2}) represent the transmission probability $\Gamma(\omega)$ for radiation to tunnel through the curvature-induced potential barrier of the Frolov black hole and reach spatial infinity. In both figures, the curves display the characteristic step-like profile of barrier scattering:
\[
\Gamma(\omega) \approx 0 \quad (\omega \ll V_{\max}^{1/2}), 
\qquad 
\Gamma(\omega) \to 1 \quad (\omega \gg V_{\max}^{1/2}),
\]
showing that low-frequency waves are strongly reflected, while high-frequency waves penetrate the barrier almost freely. 

A clear multipole hierarchy is visible in both figures. As $\ell$ (electromagnetic case) or $|k|$ (Dirac case) increases, the transmission curves shift toward higher frequencies. Higher multipole moment strengthens the centrifugal contribution to the effective potential, raising and widening the barrier and thus suppressing tunneling at a given frequency.

When comparing the two figures, the Dirac field exhibits a transmission curve that is slightly shifted toward lower frequencies relative to the electromagnetic case. In particular, for comparable multipole indices, the Dirac grey-body factors rise from near zero and approach unity at smaller values of $\omega$. This behaviour originates from the spin-dependent structure of the effective potentials: while the electromagnetic potential is directly proportional to $f(r)\ell(\ell+1)/r^2$, the Dirac potential contains additional terms involving derivatives of the metric function. As seen from the effective potential profiles discussed earlier, the peak of the Dirac potential is generally lower than that of the electromagnetic potential for similar angular indices as shown in Figs.(\ref{Pot1}) to (\ref{Pot4}). A lower barrier height allows fermionic waves to tunnel through the spacetime more easily, so the transition from reflection-dominated to transmission-dominated regimes occurs at smaller frequencies. Nevertheless, both fields preserve the same qualitative multipole hierarchy and approach the universal high-frequency limit $\Gamma \to 1$, confirming that ultraviolet modes are largely insensitive to both spin and the internal regular core structure.

\vspace{6mm}

\section{Unruh Temperature}
\label{sec:GR6}

In this section, we examine the Unruh temperature associated with the spacetime under consideration.  
This temperature reflects the surface gravity or, equivalently, the proper acceleration felt by an observer stationed at a fixed radial distance from the black hole. Such a property has also been explored in various alternative theories of gravity. The Unruh temperature can be expressed in terms of the red-shifted surface gravity, or more generally through covariant quantities involving the gravitational potential and the timelike Killing vector field. Previously, the Unruh temperature has been investigated in various setups \cite{Verlinde:2010hp, Konoplya:2010ak} 

Let $\phi$ denote the gravitational potential and $\xi^\alpha$ a timelike Killing vector.  
The potential is defined as
\begin{equation}
\phi = \frac{1}{2}\,\ln\!\bigl[-g_{\alpha\beta}\,\xi^{\alpha}\xi^{\beta}\bigr].
\end{equation}

The quantity
\begin{equation}
e^{\phi}=\sqrt{-\,g_{\alpha\beta}\,\xi^{\alpha}\xi^{\beta}}
\end{equation}
is the redshift factor associated with the static Killing field. It measures how the proper time of a stationary observer at radius $r$ relates to the coordinate time
measured by an observer at spatial infinity. In an asymptotically flat spacetime,
this factor approaches unity as $r\to\infty$, ensuring the standard normalization
of the Killing vector.

The background geometry is assumed to be a static solution that admits a global timelike Killing vector $\xi^{\alpha}$. For the static and spherically symmetric spacetime considered here, we choose the timelike Killing vector associated with time translations,
\(\xi^\mu = (\partial_t)^\mu\).
Its covariant components are given by
\(\xi_\mu = g_{\mu\nu}\xi^\nu = (-f(r),0,0,0)\),
where \(f(r)\) is the metric function.
This Killing vector is normalized such that
\(\xi^\mu \xi_\mu \to -1\) at spatial infinity.

The local acceleration then follows from
\begin{equation}
a^{\alpha} = - g^{\alpha\beta}\nabla_{\beta}\phi,
\end{equation}

where 
\begin{equation}
\phi(r) = \frac{1}{2}\ln f(r)
\end{equation}

The corresponding Unruh temperature is
\begin{equation}
T_{\text{Unruh}}
   = \frac{\hbar}{2\pi}\, e^{\phi}\, n_{\alpha}\nabla^{\alpha}\phi ,
\end{equation}
\\
Here $n^{\alpha}$ is a unit vector that is everywhere orthogonal to the timelike
Killing vector $\xi^{\alpha}$. This equation can be rewritten as \cite{Unruh:1976db}
\begin{equation}
T_{\text{Unruh}}
   = \frac{\hbar}{2\pi}\, e^{\phi}
      \sqrt{g^{\alpha\beta}\,\partial_{\alpha}\phi\,\partial_{\beta}\phi } 
\end{equation}

For a static, spherically symmetric black hole with metric function $f(r)$, a static observer at radius $r$ measures by setting $\hbar = 1$, and $M = 1$ this becomes, 

\begin{equation}
T(r) = \frac{ r \left( q^{4} \alpha^{2} + M r^{2}\left(r^{3} - 4 M \alpha^{2}\right) - 
   q^{2}\left(r^{4} + 2 M r \alpha^{2}\right) \right) }
{ 2 \pi \left( r^{4} + q^{2} \alpha^{2} + 2 M r \alpha^{2} \right)^{2} }
\end{equation}

\begin{figure}[h!]
    \centering
    \includegraphics[width=0.48\textwidth]{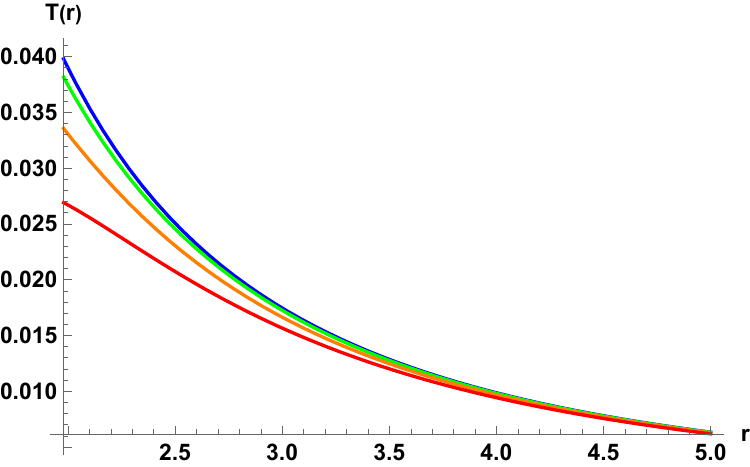}
    \caption{Unruh temperature as a function of the radial coordinate for $M = 1$ and fixed $q = 0.2$
with $\alpha_0 = 0$ (blue), $\alpha_0 = 0.2$ (green), $\alpha_0 = 0.4$ (orange), and $\alpha_0 = 0.6$ (red).}

    \label{unruh1}
\end{figure}

\begin{figure}[h!]
    \centering
    \includegraphics[width=0.52\textwidth]{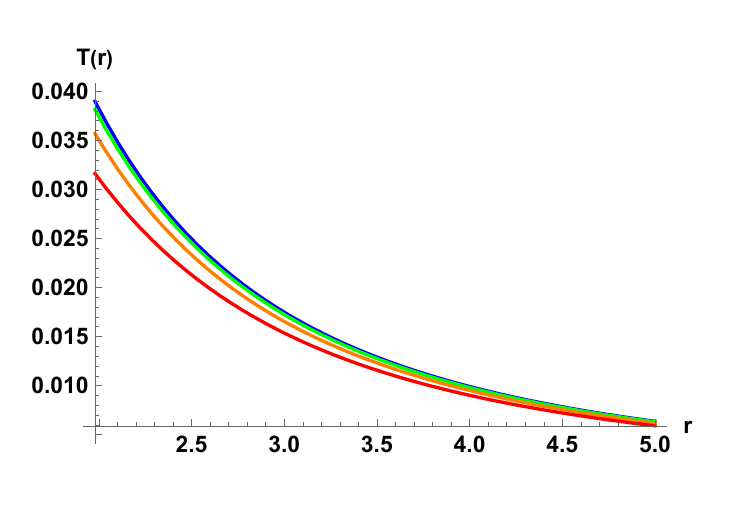}
    \caption{Unruh temperature as a function of the radial coordinate for $M = 1$ and fixed $\alpha_0 = 0.2$
with $q = 0$ (blue), $q = 0.2$ (green), $q = 0.4$ (orange) and $q = 0.6$ (red).}

    \label{unruh2}
\end{figure}

Figures (\ref{unruh1}) and (\ref{unruh2}) depict the variation of the Unruh--Verlinde temperature $T(r)$ with respect to the radial coordinate $r$ for the Frolov black hole. In both cases, the temperature exhibits a monotonically decreasing behavior with increasing $r$, approaching zero asymptotically. This trend is consistent with the expectation that the effective temperature perceived by a static observer decreases with distance from the gravitational source due to the redshift of local acceleration.

In Fig.(\ref{unruh1}), the BH parameter $\alpha_0$ is varied while keeping the charge fixed. Similar to the effect of $q$, increasing $\alpha_0$ also lowers the Unruh temperature profile. It can be seen that a larger $\alpha_0$ weakens the spacetime curvature near the core, reducing the surface gravity at the horizon and hence the temperature. This behavior indicates that the inclusion of parameter effects leads to a more stable and colder black hole configuration.\\
In Fig.(\ref{unruh2}), the charge parameter $q$ is varied while keeping the regularization parameter $\alpha_0 = 0.2$ fixed. It can be seen that increasing $q$ leads to a systematic decrease in the Unruh temperature across the entire radial domain. The presence of electric charge effectively weakens the gravitational attraction experienced by a static observer, resulting in a reduced proper acceleration and, consequently, a lower Unruh temperature. The curves clearly demonstrate that the black hole becomes thermodynamically "cooler" as the charge grows, approaching the extremal limit where $T \rightarrow 0$.

Both parameters $q$ and $\alpha_0$ act to suppress the Unruh--Verlinde temperature. The temperature profiles confirm that the Frolov black hole smoothly interpolates between the classical Schwarzschild case ($q = 0$, $\alpha_0 = 0$) and a regularized, cooler configuration where parameter corrections and charge effects dominate.

\section{Discussion and Conclusions}
\label{sec:GR}


We have carried out a detailed study of the quasinormal modes (including the first overtone), grey-body factors, and Unruh temperature for test electromagnetic and Dirac fields propagating in the background of a static charged regular black hole, also known as the Frolov BH, arising in Quantum Gravity scenarios. Using the sixth-order WKB approximation for the electromagnetic field and for the Dirac field we cross-checked with higher-order approximations and Padé Averaging for numerical consistency and derived and solved the corresponding wave equations for both cases.  

The QNM spectra of the Frolov black hole, obtained in this case, reveal distinct features that set this regularized geometry apart from the classical Schwarzschild and Reissner–Nordström (RN) spacetimes. For all modes considered, the real part of the frequency $Re(\omega)$ increases with both the electric charge $q$ and the BH parameter $\alpha_0$, indicating that the inclusion of these parameters strengthens the effective potential barrier and raises the oscillation frequency of perturbations. Simultaneously, the magnitude of the imaginary part $|Im(\omega)|$ decreases, showing that damping becomes weaker and the perturbations persist longer.

The excellent agreement between the sixth and eighth-order WKB–Padé averaged results confirms the reliability of the computed spectra for both fundamental and overtone modes. The standard multipole and overtone hierarchies are preserved—higher $l$ (or $|k|$) leads to larger $Re(\omega)$ and stronger damping, while higher $n$ corresponds to faster decay which shows that the fundamental structure of the perturbative response remains intact. One can infer clearly from the tables how the spectra differs from the standard results of the Schwarzschild ($q = 0$ and $\alpha_0 = 0$) as well as the Reissner-Nordström case ($\alpha_0 = 0$). These findings establish that the Frolov spacetime suggest that its ringdown signature could provide a distinct observational imprint of quantum-gravity inspired regular black holes. The grey-body factors, which determine the transmission probability of Hawking radiation through the effective potential barrier, are also affected by the corrections. We find that quantum effects slightly suppress the transmission coefficients at lower frequencies, effectively modifying the spectrum of radiation that reaches a distant observer. This suppression reflects how the modified spacetime geometry alters the scattering properties of the black hole potential.  

In addition, our results indicate that the Unruh temperature decreases monotonically with increasing $\alpha_0$ while fixing the charge $q$ and vice versa, suggesting a direct influence of these corrections on the thermal properties of the horizon. These findings highlight that while QNM ringing remains a robust observable for probing strong-field regimes and constraining possible quantum gravity effects. Our analysis opens up several avenues for future investigation. It would be particularly interesting to extend this work by studying the gravitational perturbations in greater detail, as such modes are directly linked to the quasinormal spectra observable in future gravitational wave detections. A comprehensive analysis incorporating both axial and polar perturbations would help to establish the complete dynamical stability of the Frolov black hole spacetime. Furthermore, it would be valuable to explore the Hawking radiation spectra associated with this geometry, as well as to investigate its implications in the context of modified theories of gravity and quantum gravitational corrections.

\section*{Acknowledgments}

The author would like to thank R. A. Konoplya for stimulating discussions and the anonymous referees for their valuable comments, which significantly improved the manuscript. The author also expresses his gratitude to Pankaj S. Joshi and Parth Bambhaniya for giving the opportunity to visit ICSC, Ahmedabad University. 



\begin{thebibliography}{999}


\bibitem{Penrose:1964wq}
R.~Penrose,
\href{https://doi.org/10.1103/PhysRevLett.14.57}{Phys. Rev. Lett. \textbf{14}, 57-59 (1965)}



\bibitem{Hawking:1966vg}
S.~W.~Hawking,
\href{https://doi.org/10.1103/PhysRevLett.17.444}{Phys. Rev. Lett. \textbf{17}, 444-445 (1966)}



\bibitem{refId0}
Valeri P. Frolov.
\newblock Remarks on non-singular black holes.
\newblock {\em EPJ Web Conf.}, 168:01001, 2018.
\newblock doi: \href{https://doi.org/10.1051/epjconf/201816801001}{10.1051/epjconf/201816801001}.

\bibitem{Ayon-Beato:2000mjt}
E.~Ayon-Beato and A.~Garcia,
\href{https://doi.org/10.1016/S0370-2693(00)01125-4}{Phys. Lett. B \textbf{493}, 149-152 (2000)}
[arXiv:gr-qc/0009077 [gr-qc]].

\bibitem{Dymnikova:2004zc}
I.~Dymnikova,
\href{https://doi.org/10.1088/0264-9381/21/18/009}{Class. Quant. Grav. \textbf{21}, 4417-4429 (2004)}
[arXiv:gr-qc/0407072 [gr-qc]].

\bibitem{Dymnikova1992}
I. Dymnikova.
\newblock Vacuum nonsingular black hole.
\newblock {\em General Relativity and Gravitation}, 24:235–242, 1992.
\newblock doi: \href{https://doi.org/10.1007/BF00760226}{10.1007/BF00760226}.

\bibitem{Ansoldi:2008jw}
S.~Ansoldi,
\href{https://arxiv.org/abs/0802.0330}
{[arXiv:0802.0330 [gr-qc]]}.

\bibitem{Ovalle:2024wtv}
J.~Ovalle,
\href{https://doi.org/10.1103/PhysRevD.109.104032}{Phys. Rev. D \textbf{109}, no.10, 104032 (2024)}
[arXiv:2405.06731 [gr-qc]].

\bibitem{Ovalle:2025pue}
J.~Ovalle,
\href{https://arxiv.org/abs/2509.00816}[arXiv:2509.00816 [gr-qc]].

\bibitem{Lemos:2011dq}
J.~P.~S.~Lemos and V.~T.~Zanchin,
\href{https://doi.org/10.1103/PhysRevD.83.124005}{Phys. Rev. D \textbf{83}, 124005 (2011)}
[arXiv:1104.4790 [gr-qc]].

\bibitem{Hayward:2005gi}
S.~A.~Hayward,
\href{https://doi.org/10.1103/PhysRevLett.96.031103}{Phys. Rev. Lett. \textbf{96}, 031103 (2006)}
[arXiv:gr-qc/0506126 [gr-qc]].

\bibitem{Frolov:2016pav}
V.~P.~Frolov,
\href{https://doi.org/10.1103/PhysRevD.94.104056}{Phys. Rev. D \textbf{94}, no.10, 104056 (2016)}
[arXiv:1609.01758 [gr-qc]].

\bibitem{Xiang:2013sza}
L.~Xiang, Y.~Ling and Y.~G.~Shen,
\href{https://doi.org/10.1142/S0218271813420169}{Int. J. Mod. Phys. D \textbf{22}, 1342016 (2013)}
[arXiv:1305.3851 [gr-qc]].

\bibitem{Culetu:2013fsa}
H.~Culetu,
\href{https://arxiv.org/abs/1305.5964}{[arXiv:1305.5964 [gr-qc]]}.

\bibitem{Culetu:2014lca}
H.~Culetu,
\href{https://doi.org/10.1007/s10773-015-2521-6}{Int. J. Theor. Phys. \textbf{54}, no.8, 2855-2863 (2015)}
[arXiv:1408.3334 [gr-qc]].

\bibitem{Rodrigues:2015ayd}
M.~E.~Rodrigues, E.~L.~B.~Junior, G.~T.~Marques and V.~T.~Zanchin,
\href{https://doi.org/10.1103/PhysRevD.94.024062}{Phys. Rev. D \textbf{94}, no.2, 024062 (2016)}
[arXiv:1511.00569 [gr-qc]].

\bibitem{Simpson:2019mud}
A.~Simpson and M.~Visser,
\href{https://doi.org/10.3390/universe6010008}{Universe \textbf{6}, no.1, 8 (2019)}
[arXiv:1911.01020 [gr-qc]].

\bibitem{Ghosh:2014pba}
S.~G.~Ghosh,
\href{https://doi.org/10.1140/epjc/s10052-015-3740-y}{Eur. Phys. J. C \textbf{75}, no.11, 532 (2015)}
[arXiv:1408.5668 [gr-qc]].

\bibitem{Li:2016yfd}
X.~Li, Y.~Ling, Y.~G.~Shen, C.~Z.~Liu, H.~S.~He and L.~F.~Xu,
\href{https://doi.org/10.1016/j.aop.2018.07.021}{Annals Phys. \textbf{396}, 334-350 (2018)}
[arXiv:1611.09016 [gr-qc]].

\bibitem{Martinis:2010zk}
M.~Martinis and N.~Perkovic,
\href{https://arxiv.org/abs/1009.6017}
{[arXiv:1009.6017 [gr-qc]]}.

\bibitem{Ling:2021olm}
Y.~Ling and M.~H.~Wu,
\href{https://doi.org/10.1088/1361-6382/acc0c9}{Class. Quant. Grav. \textbf{40}, no.7, 075009 (2023)}
[arXiv:2109.05974 [gr-qc]].







\bibitem{Konoplya:2011qq}
R.~A.~Konoplya and A.~Zhidenko,
\href{https://doi.org/10.1103/RevModPhys.83.793}{Rev. Mod. Phys. \textbf{83}, 793-836 (2011)}
[arXiv:1102.4014 [gr-qc]].

\bibitem{Berti:2009kk}
E.~Berti, V.~Cardoso and A.~O.~Starinets,
\href{https://doi.org/10.1088/0264-9381/26/16/163001}{Class. Quant. Grav. \textbf{26}, 163001 (2009)}
[arXiv:0905.2975 [gr-qc]].

\bibitem{Konoplya:2024lir}
R.~A.~Konoplya and A.~Zhidenko,
\href{https://doi.org/10.1088/1475-7516/2024/09/068}{JCAP \textbf{09}, 068 (2024)}
[arXiv:2406.11694 [gr-qc]].

\bibitem{Tang:2025mkk}
C.~Tang, Y.~Ling and Q.~Q.~Jiang,
"Correspondence between grey-body factors and quasinormal modes for regular black holes with sub-Planckian curvature,''
\href{https://inspirehep.net/literature/2905022}.

\bibitem{Bolokhov:2024otn}
S.~V.~Bolokhov and M.~Skvortsova,
\href{https://doi.org/10.1088/1475-7516/2025/04/025}{JCAP \textbf{04}, 025 (2025)}
[arXiv:2412.11166 [gr-qc]].

\bibitem{Dubinsky:2025fwv}
A.~Dubinsky,
\href{https://doi.org/10.53941/ijgtp.2025.100002}{International Journal of Gravitation and Theoretical Physics. \textbf{1}, 2 (2025)}

\bibitem{Konoplya:2021ube}
R.~A.~Konoplya,
\href{https://doi.org/10.1016/j.physletb.2021.136734}{Phys. Lett. B \textbf{823}, 136734 (2021)}
[arXiv:2109.01640 [gr-qc]].

\bibitem{Song:2024kkx}
Z.~Song, H.~Gong, H.~L.~Li, G.~Fu, L.~G.~Zhu and J.~P.~Wu,
\href{https://doi.org/10.1088/1572-9494/ad5717}{Commun. Theor. Phys. \textbf{76}, no.10, 105401 (2024)}
[arXiv:2406.04787 [gr-qc]].

\bibitem{Lopez:2018aec}
L.~A.~Lopez and V.~Hinojosa,
\href{https://doi.org/10.1088/1361-6382/acc0c9}{Can. J. Phys. \textbf{99}, no.1, 44-48 (2021)}
[arXiv:1810.09034 [gr-qc]].



\bibitem{LISA:2022kgy}
K.~G.~Arun \textit{et al.} [LISA],
\href{https://doi.org/10.1007/s41114-022-00036-9}{Living Rev. Rel. \textbf{25}, no.1, 4 (2022)}
[arXiv:2205.01597 [gr-qc]].

\bibitem{Barausse:2020rsu}
E.~Barausse, E.~Berti, T.~Hertog, S.~A.~Hughes, P.~Jetzer, P.~Pani, T.~P.~Sotiriou, N.~Tamanini, H.~Witek and K.~Yagi, \textit{et al.}
\href{https://doi.org/10.1007/s10714-020-02691-1}{Gen. Rel. Grav. \textbf{52}, no.8, 81 (2020)}
[arXiv:2001.09793 [gr-qc]].


\bibitem{DeWitt:1973uma}
C.~DeWitt and B.~S.~DeWitt, R. Ruffini, in Black Hole: les Astres Occlus (Gordon and Breachm New York, 1973)
Proceedings, Ecole d'Et{\'e} de Physique Th{\'e}orique: Les Astres Occlus: Les Houches, France, August, 1972,''


\bibitem{Li:2013fka}
J.~Li, M.~Hong and K.~Lin,
\href{https://doi.org/10.1103/PhysRevD.88.064001}{Phys. Rev. D \textbf{88}, 064001 (2013)}
[arXiv:1308.6499 [gr-qc]].

\bibitem{Kokkotas:1999bd}
K.~D.~Kokkotas and B.~G.~Schmidt,
\href{https://doi.org/10.12942/lrr-1999-2}{Living Rev. Rel. \textbf{2}, 2 (1999)}
[arXiv:gr-qc/9909058 [gr-qc]].


\bibitem{Schutz_1985}
B.~F.~Schutz and C.~M.~Will,  
  \href{https://doi.org/10.1086/184453}{APJL \textbf{291}, L33–L36, (1985).}

\bibitem{Iyer_1987}
S.~Iyer and C.~M.~Will,  
\href{https://doi.org/10.1103/PhysRevD.35.3621}{Phys. Rev. D \textbf{35}, 3621–3631, (1987).}.


\bibitem{Konoplya:2003ii}
R.~A.~Konoplya,
\href{https://doi.org/10.1103/PhysRevD.68.024018}{Phys. Rev. D \textbf{68}, 024018 (2003)}
[arXiv:gr-qc/0303052 [gr-qc]].


\bibitem{Matyjasek:2017psv}
J.~Matyjasek and M.~Opala,
\href{https://doi.org/10.1103/PhysRevD.96.024011}{Phys. Rev. D \textbf{96}, no.2, 024011 (2017)}
[arXiv:1704.00361 [gr-qc]].


\bibitem{Hatsuda:2019eoj}
Y.~Hatsuda,
\href{https://doi.org/10.1103/PhysRevD.101.024008}{Phys. Rev. D \textbf{101}, no.2, 024008 (2020)}
[arXiv:1906.07232 [gr-qc]].

\bibitem{Konoplya:2019hlu}
R.~A.~Konoplya, A.~Zhidenko and A.~F.~Zinhailo,
\href{https://doi.org/10.1088/1361-6382/ab2e25}{Class. Quant. Grav. \textbf{36} (2019), 155002}
[arXiv:1904.10333 [gr-qc]].

\bibitem{Matyjasek:2019eeu}
J.~Matyjasek and M.~Telecka,
\href{https://doi.org/10.1103/PhysRevD.100.124006}{Phys. Rev. D \textbf{100}, no.12, 124006 (2019)}
[arXiv:1908.09389 [gr-qc]].

\bibitem{Konoplya202X}
R.~Konoplya, ``Mathematica R package for QNM and grey-body factor calculations,'' available at: \url{https://goo.gl/nykYGL}.


\bibitem{Page:1976df}
D.~N.~Page,
\href{https://doi.org/10.1103/PhysRevD.13.198}{Phys. Rev. D \textbf{13}, 198-206 (1976)}

\bibitem{Konoplya:2019ppy}
R.~A.~Konoplya and A.~F.~Zinhailo,
\href{https://doi.org/10.1103/PhysRevD.99.104060}{Phys. Rev. D \textbf{99}, no.10, 104060 (2019)}
[arXiv:1904.05341 [gr-qc]].



\bibitem{Verlinde:2010hp}
E.~P.~Verlinde,
\href{https://doi.org/10.1007/JHEP04(2011)029}{JHEP \textbf{04}, 029 (2011)}
[arXiv:1001.0785 [hep-th]].

\bibitem{Konoplya:2010ak}
R.~A.~Konoplya,
\href{https://doi.org/10.1140/epjc/s10052-010-1424-1}{Eur. Phys. J. C \textbf{69}, 555-562 (2010)}
[arXiv:1002.2818 [hep-th]].

\bibitem{Unruh:1976db}
W.~G.~Unruh,
\href{https://doi.org/10.1103/PhysRevD.14.870}{Phys. Rev. D \textbf{14}, 870 (1976)}



\end{thebibliography}
\end{document}